\documentclass[12pt]{article}

\usepackage{arxiv}

\usepackage[utf8]{inputenc} 
\usepackage[T1]{fontenc}    
\usepackage{hyperref}       
\usepackage{url}            
\usepackage{booktabs}      
\usepackage{amsfonts}      
\usepackage{nicefrac}       
\usepackage{microtype}      
\usepackage{lipsum}
\usepackage{graphicx}
\usepackage{subcaption}
\usepackage[export]{adjustbox}
\usepackage{chngcntr}
\graphicspath{ {./images/} }
\usepackage{natbib}
\usepackage{float}
\usepackage{longtable}
\setcitestyle{authoryear,open={(},close={)}}
\usepackage{setspace}
\usepackage{adjustbox}
\usepackage{rotating}
\usepackage[hang,flushmargin]{footmisc}
\usepackage{enumitem}
\usepackage{ulem}
\usepackage{setspace}
\usepackage{silence}
\renewcommand*\url[1]{\href{http://#1}{\texttt{#1}}}

\usepackage{color}

\title{Bending the Automation Bias Curve: A Study of Human and AI-based Decision Making in National Security Contexts}
\author{ 
    Michael C. Horowitz \\
    University of Pennsylvania\\
   \AND
   Lauren Kahn\\
   Council on Foreign Relations\\
   }

\begin{document}

\maketitle


\clearpage

\begin{abstract}
Uses of artificial intelligence (AI), especially those powered by machine learning approaches, are growing in sectors and societies around the world. How will AI adoption proceed, especially in the international security realm? Research on automation bias suggests that humans can often be overconfident in AI, whereas research on algorithm aversion shows that, as the stakes of a decision rise, humans become more cautious about trusting algorithms. We theorize about the relationship between background knowledge about AI, trust in AI, and how these interact with other factors to influence the probability of automation bias in the international security context. We test these in a preregistered task identification experiment across a representative sample of 9000 adults in 9 countries with varying levels of AI industries. The results strongly support the theory, especially concerning AI background knowledge. A version of the Dunning Kruger effect appears to be at play, whereby those with the lowest level of experience with AI are slightly more likely to be algorithm-averse, then automation bias occurs at lower levels of knowledge before leveling off as a respondent's AI background reaches the highest levels. Additional results show effects from the task's difficulty, overall AI trust, and whether a human or AI decision aid is described as highly competent or less competent.
\end{abstract}


{Word Count: 6900}

\setstretch{2}
\clearpage

\section{Introduction}

The integration of advances in artificial intelligence (AI) by governments worldwide raises significant questions for politics and society. The delegation of high-level tasks to machines raises the prospect of accidents and could generate challenges for accountability, especially in high-stakes international security contexts such as international crises. Complicating these challenges is the potential for automation bias. Automation bias is the “tendency [for human operators] to over-rely on automation” \citep{Goddard2012,SkitkaMosiere1999} and automated systems by which these systems and their outputs become a “heuristic replacement of vigilant information seeking and processing” \citep{MosierSkitka1996}. Given the growing interest by militaries and other national security institutions in adopting AI and autonomous systems, human-machine relationships are increasingly central to life and death issues. What do these trends mean for international relations, since many international relations theories are predicated on the role of human agency in making decisions and how human-constructed politics---individual, domestic, and international ---shape international politics? 

How increasing reliance on AI will influence decision-making in international politics is undertheorized. We theorize that key drivers of automation bias for AI systems in national security contexts are experiential and attitudinal. Environmental factors such as task difficulty, background knowledge---familiarity, knowledge, experience---with AI, trust and confidence in AI, as well as self-confidence determine how much humans rely on machines---or other humans---when making decisions. In particular, we theorize that a version of the Dunning-Kruger effect is at play, where algorithm aversion is highest at the lowest levels of knowledge, flips to automation bias at low levels of knowledge, then levels off at high levels of knowledge.

Existing evidence about automation bias and algorithm aversion tends to feature relatively small sample sizes restricted to the United States. Automation bias experiments also rarely focus on a prominent use case for AI-enabled autonomous systems---military uses. To fill this gap, we test our hypotheses using a novel, pre-registered scenario-based survey experiment with 9,000 respondents across nine countries to evaluate the conditions in which automation bias is more likely, using national security scenarios that mirror current and likely future uses of AI algorithms.\footnote{preregistration link will be inserted upon article acceptance. As described below, we also choose a specific scenario where there is no reason to expect that the attitudes of those who would be making judgments in the real world would substantially differ from the general public.}

The results show the powerful impact of background knowledge, experience, and familiarity with AI on the subsequent use of AI. As hypothesized, a version of the Dunning-Kruger effect is at play. Those with the lowest level of experience with AI are slightly more likely to be algorithm-averse, then automation bias occurs at lower levels of knowledge before leveling off as a respondent's AI background reaches the highest levels. Additional results show effects from the task's difficulty, overall AI trust, and whether a human or AI decision aid is described as highly competent or less competent.

The experiment and our findings contribute to the international relations literature in three ways. First, the results give us information on how AI is likely to shape how people make decisions about key areas of international politics, so it sheds light on decision-making processes at the core of international relations debates. Second, the focus on AI and automation bias bolsters growing research on how emerging technologies shape international politics \citep{horowitz2020emerging,KrepsDrones,Sechser2019,Schmidt2022,JJ2020}. Third, this research contributes to ongoing debates about the role of trust and confidence in political decision-making, which is particularly salient now in the context of growing interest in generative AI and large language models such as ChatGPT and GPT-4.

Moreover, the findings can also contribute to ongoing public policy debates about AI and automation. Militaries worldwide seek to integrate and leverage AI capabilities into existing systems and processes, sometimes with fatal consequences. For example, in 2003, two separate accidents were caused, in part, by failures in a Patriot missile’s automated tracking and Identification Friend or Foe(IFF) systems---the system which allows the missiles to differentiate between a friendly aircraft and enemy anti-radiation or ballistic missiles, and between enemy aircraft---resulted in three fatalities. These accidents represented a complicated cascade of simultaneous human and machine failures, precipitated by an error in an automated decision aid and facilitated by established organizational practice that defaulted towards the automated settings \citep{PatriotWars, Atherton2022, AccidentReport}.

Thus, in addition to the importance for international relations, the results can inform the development of guidelines, AI education, and training programs that can improve decision-making in human-AI teams, mitigate the risk of accidents and failures, and ultimately enhance the safety and effectiveness of these systems. In what follows, we lay out our theory and hypotheses, describing the importance for international politics. We then introduce the survey experiment and overall research design, describe the results, and then conclude with a discussion of limitations and next steps.

\section{Theory}

There is growing interest in how technological change surrounding robotics, autonomous systems, and artificial intelligence and machine learning, will influence international politics \cite{hudson2019artificial, horowitz2018artificial, horowitz2020emerging, jensen2020algorithms, johnson2021artificial}. Existing work tends to focus on questions surrounding drone strikes \cite{lin2022wargame, mir2019drones, johnston2016impact, kreps2016international, horowitz2016separating}, uses of AI in nuclear command and control, \cite{fitzpatrick2019artificial,  johnson2021catalytic, Sechser2019}, and autonomous weapon systems \cite{horowitz2019speed, scharre2018army}.  How AI will shape decision-making as a whole in international relations remains understudied, especially outside the realm of crisis escalation \cite{horowitz2022algorithms}. How humans make choices about whether and how to use algorithms will be a critical part of that equation.

Automation bias refers to the tendency of humans to rely on AI-enabled decision aids above and beyond the extent to which they should, given the reliability of the algorithms. Algorithm aversion refers to the opposite---the tendency of humans in some situations to discard algorithms in favor of their own judgment despite evidence in favor of relying on an algorithm. Several factors may shape when and how automation bias or algorithm aversion manifests and the frequency with which it occurs. These include experiential factors such as familiarity with and knowledge of the system and like technologies, attitudinal factors such as trust and confidence, and environmental factors such as task difficulty and time constraints \citep{SouthernArnstern2009, BarkatBusuioc2022, SimmonsMassey2015, GoddardRoudsari2014}. This paper focuses primarily on the first two: experiential and attitudinal.\footnote{This paper primarily addresses the experiential and attitudinal factors, as the effects of certain environmental factors are already well theorized and researched. We do not propose any novel hypotheses regarding these factors.}

This paper focuses primarily on experiential and attitudinal factors, bringing research from other fields into political science.  The effects of certain environmental factors are already well theorized and researched, so we do not propose any novel hypotheses regarding those factors \cite{Goddard2012, TaskComplexity2007,povyakalo_alberdi_strigini_ayton_2013,lyell_coiera_2016}.


\subsection{Experiential: familiarity with and knowledge of AI}

The way knowledge and experience influence how individuals and countries behave is a long-running topic in international politics \citep{haas1992introduction, reiter1994learning}. Existing research suggests that experience will generate greater trust and reduce aversion to using algorithms, while time-constraint elements will increase the likelihood of automation bias. Knowledge, familiarity, and experience with AI could shape how individuals react to algorithms designed for new situations and choose whether to use algorithms in general. For clarity and brevity, we will use the term "background" to encompass all three distinct but related factors: knowledge, familiarity, and experience. Behavioral science work shows that while having no knowledge of technology can lead to fear and rejection, limited background in technology can lead to overconfidence in its capabilities. People with limited initial background with a wide variety of topical areas become subject to the “beginners bubble” \citep{SanchezDunning2018}, an illustration of the Dunning-Kruger effect whereby those with surface-level knowledge become overconfident in that knowledge, leading to sub-optimal decision-making \citep{KrugerDunning1999}. As people gain more experience, familiarity, and knowledge, the degree of overconfidence declines.

When it comes to assessing applications of AI and susceptibility to automation bias, the relationship between background in AI and reliance on automation should therefore be non-linear. Those with no experience, familiarity, or knowledge should be skeptical of AI, meaning they are also unlikely to be prone to automation bias. Those with limited backgrounds should be the most susceptible to automation bias because they have just enough knowledge, familiarity, and experience to think they understand AI but not enough to recognize limits and issues with applications. Finally, those with a substantial background in AI, on average, should be more evenly positioned between aversion to AI and automation bias---in theory, relying on the AI system an "appropriate" amount, in line with both its established abilities and limitations, proportionate to its expected performance and accuracy. In other words, they know enough to realize both the utility of algorithms in some cases and know when to question or check algorithmic outputs.\footnote{See the preregistration of this study for additional information.} For example, large language models have shown a propensity to generate irrelevant, nonsensical, or even false content. This phenomenon, known as hallucination, is a recent, concrete example of the importance of recognizing and rectifying automation bias \citep{metz_2023}. For instance, a large language model may list a citation for a nonexistent article yet attribute it to a legitimate author or journal (or both). Failure of the human not to confirm this information has already led to errors---StackOverflow, a public forum for programming help, already banned answers generated by ChatGPT due to the high incidences of subtle errors and factual inaccuracies \citep{vincent_2022}. 

Figure~\ref{fig:reliancefig} below illustrates a stylized take on this relationship.

\begin{figure}[H]
    \centering
    \includegraphics[width = 0.9\linewidth]{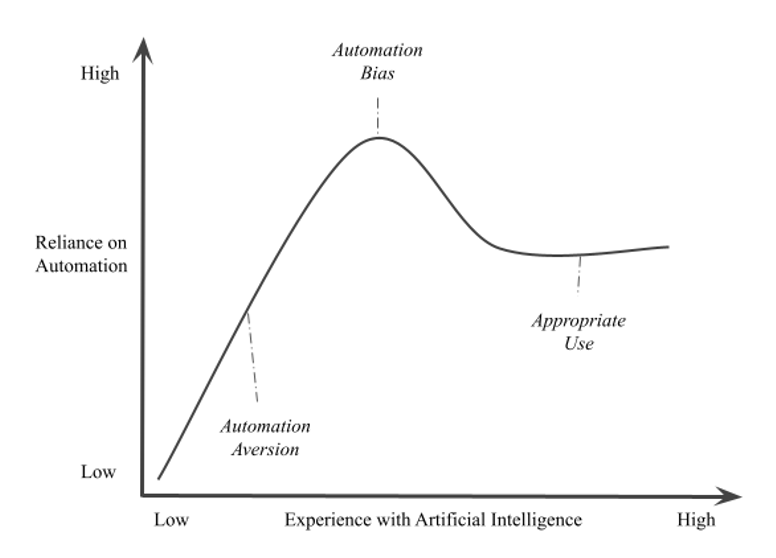}
    \caption{Reliance on automation relative to prior background in AI.}
    \label{fig:reliancefig}
\end{figure}

We theorize that background in AI (knowledge, familiarity, and experience) will influence how willing respondents are to rely on input received from AI-based systems and algorithms.

\noindent\textbf{\textit{Hypothesis 1: Those with the lowest levels of experience, knowledge, and/or familiarity (background) are relatively more averse to AI; people with middle levels of background are relatively over-reliant on AI, and those with the highest levels of background are relatively appropriately reliant on AI.}}

\subsection{Attitudinal: trust and confidence in AI}

Trust is an essential topic in international relations, making it critical to understand it in the AI context as well \citep{kydd2007trust, hoffman2002conceptualization}. In addition to firsthand knowledge of, familiarity with, and experience with AI, attitudinal factors should influence the likelihood of automation bias or algorithm aversion. Attitudinal factors measure whether individuals using AI and AI-enabled systems trust the system or algorithm to work as expected. Additionally, whether the operator has trust and/or confidence in the system to aid them in completing the task will also be relative to how much confidence and or trust they have in themselves to complete the task. 

The distinction between trust and confidence matters. Luhmann argues that trust and/or confidence influence individuals making decisions under conditions of risk \citep{Luhmann1979, Luhmann_1988}. Trust refers to a condition of individual responsibility and knowledge, where someone believes in a specific actor, often due to knowledge or the perception of shared experiences \citep{SeligmanMontgomery2019}. Trust is an active decision that involves an agenda---someone chooses to trust someone else. Confidence refers to the ability to predict the behavior of others not due to individual knowledge or experience but because of laws, social norms, or established benchmarks for success or an acceptable margin of error; the active choice is not necessarily required. Confidence derives from societal expectations about norms and laws rather than shared experience. 

For AI, one might imagine that confidence comes not from shared experience or familiarity with how algorithms work but faith in the system---the coders programming algorithms, the testers determining if they are reliable, and the evaluation processes surrounding the design, deployment, and use of algorithms as a whole. Confidence might also come from assessments of data, such as externally provided data on the reliability of an algorithm, system, or process that one does not personally understand.

The distinction between trust and confidence could play a critical role in helping explain individual and organizational decisions about AI adoption. Shared experiences and familiarity generate trust. It is not necessarily the case, however, that shared experiences and familiarity leads to more support for AI in a linear fashion because experience and knowledge could lead to knowledge of the limitations of AI.

As technology develops, there is initial hype and excitement that leads to an overestimation of the actual effectiveness of the system. As the technology develops further, the relationship reverses, and there is a “trust gap,” wherein those early expectations have been shattered by reality. Yet, while the technology continued to improve, it had not yet earned back users' trust. In the final stages, a mature technology has stabilized and reached a high level of effectiveness and familiarity, confidence, and trust in the system again, resulting in overconfidence (albeit a more stable one) in its abilities. 

This aligns with literature surrounding “hype cycles” in general and technology adoption processes, such as the Gartner Hype Cycle. In the early stages of technology development, when visibility is high, there can be a peak of inflated expectations, eventually followed by a steep drop in excitement - the "trough of disillusionment" when there is a disconnect due to misalignment between observed performance and expectations - before eventually increasing and then stabilizing as hype becomes more proportionate to actual performance as individuals become more knowledgeable about the technology.\citep{BloschFenn2018, Bahmanziarietal2016}.

Thus, at varying points in a technology’s lifecycle, trust, confidence, and willingness to rely on and defer to that technology will differ. Assessing respondents' experience and knowledge of the system will also play a role in determining the degree of trust and confidence in the system.

Studies show that many people are hesitant to trust AI, with one survey by an AI-driven sales platform (meaning it is likely biased to highlight positive views of AI) demonstrating that 42\% of US respondents did not generally trust AI \citep{Dujmovic2017}. Distrust of AI is heightened in cases where the system has already experienced a failure, even if it is normally highly reliable \citep{ValenciaBarrero2014}. Distrust in AI greatly undermines the effectiveness of AI capabilities. 

\noindent\textbf{\textit{Hypothesis 2: The more trusting of and open to AI technologies in general, the more likely they are to trust and therefore rely on, the recommendation of an AI-enabled system.}}

\noindent\textbf{\textit{Hypothesis 3: The more testing and training a decision aid (AI or human) is described to have, the more confidence the respondent will have in that aid, and the more likely they will rely on its advice.}}

Finally, another dimension of confidence is the relationship between self-confidence in the ability to do a task and willingness to switch answers in response to human or algorithm input. Those that view themselves as more competent on their own should be less likely to trust decision-making aids. Existing studies show that the relationship between self-confidence and confidence in AI shows that “human self-confidence significantly contributes to their acceptance of AI decisions” and that "humans often misattribute blame to themselves and enter a vicious cycle of relying on a poorly performing AI" \citep{Chong2022}. For example, in \cite{Chong2022}, self-confidence remained the predominant factor across all test groups: low self-confidence can increase people’s willingness to rely on AI systems. In contrast, high self-confidence can cause people to reject input from automated systems unfairly. Thus, managing the role of self-confidence in the ability to complete the task is critical in understanding the biases people have towards automated systems and AI and also towards fully grasping the decision-making process that occurs when utilizing AI input. 

\noindent\textbf{\textit{Hypothesis 4: The higher the level of self-confidence in the ability to do a task, the lower the probability that a decision aid will influence views of that task.}}

In sum, we hypothesize three attitudinal factors will impact rates of automation bias --- two relating to the system and one to the respondents themselves --- 1) trust in the system, 2) confidence in the system, and 3) self-confidence of the respondent regarding task completion. We also expect that these attitudinal factors will be directly affected by environmental factors, such as task difficulty and time constraints, that pressurize cognitive resources. However, as explained above, we don't hypothesize about these factors directly, as the literature firmly establishes these relationships.

\section{Research Design}


To test the above hypotheses, we designed a scenario-based survey experiment of the general adult public in nine countries: the United States, Russia, China, France, Australia, Japan, South Korea, Sweden, and the United Kingdom. The sample size for each country is 1,000 respondents for each, giving us a total sample size of 9,000. We obtain a representative sample of adults from each country except China and Russia, where we obtained a representative sample of urban adults.\footnote{YouGov fielded the survey in the United States and Russia, while Delta Poll fielded the survey in the remaining countries}. 

Surveys of the general public make sense for testing our theory for several reasons. First, the public can often be a good proxy for elite preferences, especially when there is no theoretical reason to think that elites would have different preferences \citep{kertzer2022experiments, kertzer2022re}. Since we are early in the age of AI and foreign policy, elites do not have decades of experience that would lead to different decision-making heuristics.

Second, we ask respondents to conduct an identification task where there is no reason to assume there are differences between the general public and those that would actually be conducting that identification in a national security situation. Third, identifying airplanes with the benefit of a decision aid is something that relatively low-level military personnel would conduct, rather than elites working in the White House. This again suggests the relevance of surveying the general public. Fourth, because we are surveying various countries, elite attitudes would be difficult to gather and unhelpful for aggregation due to distinct country-specific effects and biases for elite populations, possibly muddying results. 

\subsection{Country Selection}
Four main factors drove the selection of these countries: 1) the presence of an AI strategy, 2) each country’s national investment in AI, and 3) regional variety and prominence within international security discussions. The nine countries span the scope of Asia, Europe, and North America, including major international actors and artificial intelligence investors. Moreover, the investments made by these nine countries span the range of AI research, economic purposes, and national security investment, meaning that the countries also have a variety of national interests in the AI sphere. Finally, we also selected countries where we could reliably get an accurate and representative sample of their populations.

Each of the nine countries has published a national AI strategy outlining their national goals for using and investing in AI \citep{OECDAI2022}. The publication of national AI strategies demonstrates a prioritization of AI as an industry and technology that will make questions such as those presented in our survey potentially more relevant and potentially more recognizable to the population. 


The next element in selecting countries was the level of investment and research development each country has in AI-related fields. Stanford’s Human-Centered Artificial Intelligence (HAI) 2021 AI Index measured national AI investments across twenty-two different factors, including investment, research output, patents and intellectual property, and jobs and industries related to AI \citep{HAI2021}. In this report, the United States, China, South Korea, Australia, and the United Kingdom ranked in the top 10, while Japan, France, and Sweden ranked within the top 20 countries \citep{HAI2021}. The United States has been a prominent leader in investment in AI as most major AI industries are headquartered in the United States. While Russia did not rank as highly on the indexes or levels of investment as the other eight countries on the list, it has nonetheless made several public endeavors to invest more in AI and proclaimed its intention to become a leader in AI.

We also looked at regional variation and prominence in international security. Regional variation refers to selecting countries from around the world and those with potentially different interests and political systems. The countries selected span North America; Europe, including different regions within Europe; Asia; Eurasia, and Oceania. Additionally, they represent different political orientations within countries and different government systems, including both democracies and autocracies, and countries associated with prominent roles within broader international institutions, including the European Union, NATO, the AIIB, and the UN and UN Security Council. 

\subsection{Survey Design}

As described above, prior survey research on automation bias has often not extended to AI-enabled automation or has been limited to aviation and pilots or restricted in its sample or breadth \citep{cummings2004, Goddard2012, ParasuramanManzey2010}. Survey experiments conducted on the American public have been shown to increase our understanding of audience costs, \citep{Tomz2007, TraggerVavreck2011} the democratic peace, \citep{TomzWeeks2013} and human rights \citep{Wallace2013, Wallace2014}, while other extensive cross-national surveys highlight the large cultural differences that exist with approaches to AI \citep{Awad2018}.

Additionally, research into the efficacy of decision-making shows that decisions made by experts versus those made by the general public have little difference in their quality, meaning a general population can provide insights on elites as well \citep{Wesh2018, tetlock2009}. Therefore, the results of this survey can help us understand how decision-making in international security contexts occurs and the ability to make cross-cultural comparisons without potentially being biased by prior training that experts might have.

The experiment employs a within-participant, before-after design that evaluates how the same respondents alter their behavior based on different treatments. Participants first received instructions on the task---identifying whether an airplane belongs to their country’s military or an adversary’s military based on a set of defined characteristics. Participants then completed five practice rounds with no pressurizing constraints, time limits, or obscuring of the images, where they determined whether an airplane was an enemy or a friendly one and received live feedback on the accuracy of their identifications. The intended effect of providing live feedback was establishing a baseline for respondents regarding their effectiveness at the set task.

Following the practice rounds, participants then entered the experimental portion of the survey, where they were shown ten randomized hypothetical airplane identification scenarios with varying levels of task difficulty (the first five randomized at a lower level of difficulty, and the second five randomized at a higher level of difficulty).\footnote{In terms of difficulty, respondents had either 10 or 7 seconds to make an identification, and the airplane had either partially obscured or entirely obscured features. See the appendix for the images used and an example scenario.} Respondents were then asked to identify the aircraft in each scenario, after which they received an experimental treatment in the form of a decision aid. The decision aid was described as a team member that would provide a recommendation and would come in the form of either an AI algorithm or human analyst, with varying degrees of testing and training. The respondent would then have the opportunity to either change their answer or keep it the same as their initial identification. If the respondent was shown the control, then their initial answer was maintained as their final answer without the opportunity to switch.

The treatments were randomized across a two-by-four experimental design with a control condition for a total of nine possible treatment conditions. The decision aid could be a low-confidence human analyst or AI algorithm, a high-confidence human analyst or AI algorithm, or no suggestion at all for the control. We also randomized whether the recommendation of the decision aid was correct or incorrect. See the online appendix for specific wording. The language used to describe the human analyst and the AI algorithm and the varying levels of confidence in the system were kept identical except for whether a human or AI was providing the suggestion:

\begin{itemize}
    \item Human analyst low confidence --- correct identification
    \item Human analyst high confidence --- correct identification
    \item AI algorithm low confidence --- correct identification
    \item AI algorithm high confidence --- correct identification
    \item Human analyst low confidence --- incorrect identification
    \item Human analyst high confidence --- incorrect identification
    \item AI algorithm low confidence --- incorrect identification
    \item AI algorithm high confidence --- incorrect identification
    \item No identification suggestion
\end{itemize}

\subsubsection*{Dependent Variables}

Unless otherwise noted, the dependent variable in the analyses below is a binary variable of whether a respondent "switched" their answer after being shown a treatment (1 if switched, 0 otherwise).

\subsubsection*{Independent Variables}

We leverage the experimental design and additional questions to test the hypotheses above, including demographic data on respondents, batteries of questions about their experience with and attitudes about AI, and more. We operationalize trust in AI, confidence in the treatment, self-confidence in the ability to complete the task, cognitive pressure, and AI background as follows.

To test hypothesis 1, we introduced questions designed to measure participants’ experience with, knowledge of, and familiarity with AI and automated systems. These questions include ones designed to test their factual knowledge of AI and AI technologies, such as if they have a background in AI or computer science more broadly, were able to identify technologies that used AI, and if they can correctly answer questions about artificial intelligence technologies. Another set of questions assesses experience with AI, such as whether individuals had used AI-based systems in the past, in either work or home contexts. Finally, we gauge overall familiarity with AI with a battery of general interest, awareness, and exposure to the concepts of AI in the news, from colleagues or friends, or other sources. Additional details are in the online appendix.

These questions, in combination, allow us to create a standardized measure of AI background, thus enabling us to more reliably understand how previous interactions with or knowledge of AI influence how people interact with automated systems and whether the depth of that exposure plays a role factor. Moreover, the questions also reflect the commonly accepted styles and subject matters for measuring these indicators, making the results of this study more comparable with previous studies done previously \citep{ZhangDafoe2019}. This comparability allows this study to build upon current understandings of human interactions with AI and automated systems.

\begin{itemize}
    \item \textit{Overall AI Knowledge}: measured using how well respondents performed on two AI knowledge quiz questions.
    \item \textit{Overall AI Familiarity}: measured using a scale of how familiar respondents were with AI, based on if they had heard about AI from friends or family, read about AI in the news, from other avenues, or not at all.
    \item \textit{Overall AI Experience}: measured using two indicators, 1) whether an individual had some degree of coding or programming experience, and 2) whether a respondent had used a specific application and if they viewed that application to utilize AI technology.
    \item \textit{AI Background Index}: overall knowledge, experience, and familiarity with AI, taken together, in the form of a normalized index. Knowledge was attributed a smaller weight of 1/5, while experience and familiarity were each weighted at 2/5. This was because the knowledge questions were difficult for respondents, with only 25 respondents out of the total 9,016 getting both questions 100\% 
\end{itemize}

Figure~\ref{fig:AIBackgroundDetails} in the appendix plots the mean values for each variable. We test hypotheses 2-4 with the following variables.

\begin{itemize}
    \item \textit{Trust in AI}: To test hypothesis 2, we created an “AI Beliefs” battery of questions that can serve as a baseline measure for how trusting an individual respondent is of AI and AI-enabled systems. Our battery was derived from the top three factor-loaded items from each positive and negative sub-scale from an AI-specific scale based on a validated index that maps attitudes towards technologies, the Technology Readiness Index 2.0. \citep{Parasuraman2014, Schepman2020, Lam2008}.\footnote{The full list of questions is viewable in the appendix.} 
    \item \textit{Treatment Confidence}: To test hypothesis 3, confidence in the decision aid was set by the above language for each treatment, with high and low confidence language held constant across the human analysts and AI algorithm treatments. High confidence language was that either the analyst or AI algorithm "has undergone extensive testing and training to identify airplanes under these conditions." In contrast, the low-confidence language described the decision support aid as "still undergoing testing and being trained to identify airplanes under these conditions."
    \item \textit{Self-Confidence}: To test hypothesis 4, we operationalize the relationship between self-confidence and answer switching by examining how the number of correct identifications by participants in the practice rounds influences the probability of switching answers during the genuine identification rounds. The live feedback element of the practice rounds provides the respondent with a benchmark for how well they could complete the task, absent any pressurizing cognitive constraints.
\end{itemize}

Below, we show the summary statistics and correlation matrix for key variables in Table~\ref{Tab:SummaryStats} and appendix Figure~\ref{fig:CorrelationPlot}. As described below, since respondents each completed ten identification rounds, we present the statistics here showing the table disaggregated by identification, as opposed to grouped by respondent.

\begin{table}[!htbp] \centering 
\begin{tabular}{@{\extracolsep{5pt}}lccccccc} 
\\[-1.8ex]\hline \\[-1.8ex] 
Statistic & \multicolumn{1}{c}{N} & \multicolumn{1}{c}{Mean} & \multicolumn{1}{c}{St. Dev.} & \multicolumn{1}{c}{Min} & \multicolumn{1}{c}{Max} \\ 
\hline \\[-1.8ex] 
Sex (Female = 1) & 89,810 & 0.52 & 0.50 & 0.00 & 1.00 \\ 
Age & 90,160 & 46.51 & 16.56 & 18 & 93 \\ 
Political Ideology (Right = 10) & 70,220 & 5.61 & 2.51 & 0.00 & 10.00 \\ 
Highest Level of Education & 90,160 & 2.24 & 1.56 & 0 & 5 \\ 
Total Number of Practice Rounds Correct & 90,160 & 2.65 & 1.21 & 0 & 5 \\ 
Recieved a High Confidence Treatment & 90,160 & 0.45 & 0.50 & 0 & 1 \\ 
Recieved a Low Confidence Treatment & 90,160 & 0.44 & 0.50 & 0 & 1 \\ 
Recieved an AI Algorithm Treatment & 90,160 & 0.44 & 0.50 & 0 & 1 \\ 
Recieved a Human Analyst Treatment & 90,160 & 0.45 & 0.50 & 0 & 1 \\ 
Switched Identification After Treatment & 80,158 & 0.23 & 0.42 & 0.00 & 1.00 \\ 
AI Background Index & 90,160 & 0.23 & 0.15 & 0.00 & 0.86 \\ 
Trust in AI & 73,990 & 16.05 & 4.31 & 0.00 & 28.00 \\ 

\end{tabular}
\caption{Summary Statistics}
\label{Tab:SummaryStats}
\end{table}

It is necessary to acknowledge the complex relationships between these different variables. We theorize above that experience, familiarity, knowledge, trust, and confidence are closely related but distinct factors that may influence rates of automation bias. Our analysis reveals a positive, statistically significant correlation between background and trust in AI, demonstrating that attitudinal factors may be an intervening variable in the relationship between experiential factors and rates of automation bias. We attempted to separate these factors in our analysis to gain a more nuanced understanding of the interplay between these variables. However, there may be some degree of interdependence between these factors, which we attempt to disentangle below as much as possible.

\section{Results}

We start by evaluating the "rate of switching," the rate at which respondents opted to change their initial identification after being presented with the treatment condition in each round, our main dependent variable. Figure~\ref{fig:MeanSwitchingRateOverall} below illustrates the mean switching rate for respondents that received each treatment: high or low-confidence human analyst or AI algorithm decision support. Overall, individuals were more likely to switch to high-confidence treatments. At the high-confidence levels, individuals switched most frequently when it was a human analyst treatment, while at low-confidence levels, individuals were likelier to switch if it was an AI algorithm treatment. This suggests that the general public has different thresholds for error tolerance and expected performance for humans and AI systems. When the AI or human analyst was described as having undergone "extensive testing and training," a human elicits the most confidence. However, the AI or human analyst was described as "still undergoing testing and training;" an individual might view any AI, even an incompletely trained and tested AI algorithm, as preferable to a human analyst who might be very susceptible to human error and mistakes.

\begin{figure}[H]
    \centering
    \includegraphics[width = \linewidth]{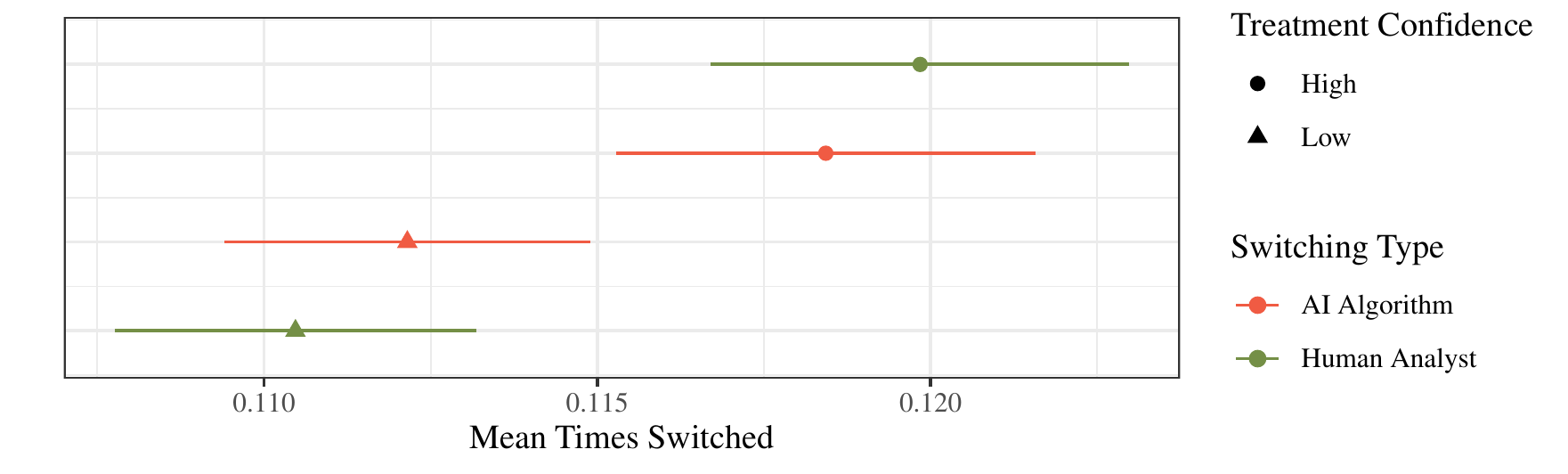}
    \caption{Mean Level of Switching Per Treatment Type}
    \label{fig:MeanSwitchingRateOverall}
\end{figure}

Table~\ref{AnalysisOfRespondentSwitching} below presents the results of an initial regression analysis designed to further test our hypotheses in models 1 and 2. The dependent variable is the same switching variable described above. For simplicity of display and interpretation, we use OLS even though the dependent variable is binary. The results are consistent, using logit models. The universe of cases is one observation per respondent identification, so there are ten observations (reflecting the ten game rounds) total per respondent. We, therefore, also cluster standard errors on the respondent. We include country-fixed effects (with the United States as a base excluded country) to account for country effects.

\begin{sidewaystable}
  \centering 
  \def\sym#1{\ifmmode^{#1}\else\(^{#1}\)\fi}
  \resizebox{\textwidth}{!}{%
    \begin{tabular}{l*{7}{c}}
                          &\multicolumn{1}{c}{(1)}&\multicolumn{1}{c}{(2)}&\multicolumn{1}{c}{(3)}&\multicolumn{1}{c}{(4)}&\multicolumn{1}{c}{(5)}&\multicolumn{1}{c}{(6)}&\multicolumn{1}{c}{(7)}\\
                    &\multicolumn{1}{c}{\shortstack{Overall\\Binary\\b/SE}}&\multicolumn{1}{c}{\shortstack{AI Condition\\Binary\\b/SE}}&\multicolumn{1}{c}{\shortstack{Human Condition\\Binary\\b/SE}}&\multicolumn{1}{c}{\shortstack{Switching in AI Condition\\AI Familiarity Model\\b/SE}}&\multicolumn{1}{c}{\shortstack{Switching in AI Condition\\AI Knowledge Model\\b/SE}}&\multicolumn{1}{c}{\shortstack{Switching in AI Condition\\AI Experience Model\\b/SE}}&\multicolumn{1}{c}{\shortstack{Switching in AI Condition\\AI Background Model\\b/SE}}\\
\hline
Political Ideology  &                     &                     &                     &       0.032\sym{***}&       0.037\sym{***}&       0.039\sym{***}&       0.034\sym{***}\\
                    &                     &                     &                     &     (0.010)         &     (0.010)         &     (0.010)         &     (0.010)         \\
Level of Difficulty &       0.034\sym{***}&       0.035\sym{***}&       0.041\sym{***}&       0.200\sym{***}&       0.200\sym{***}&       0.201\sym{***}&       0.200\sym{***}\\
                    &     (0.003)         &     (0.004)         &     (0.004)         &     (0.034)         &     (0.034)         &     (0.034)         &     (0.034)         \\
Practice Round Accuracy&      -0.008\sym{***}&      -0.011\sym{***}&      -0.008\sym{***}&      -0.042\sym{**} &      -0.047\sym{**} &      -0.047\sym{**} &      -0.044\sym{**} \\
                    &     (0.002)         &     (0.002)         &     (0.002)         &     (0.019)         &     (0.019)         &     (0.019)         &     (0.019)         \\
Age                 &      -0.000\sym{*}  &      -0.000\sym{**} &      -0.000         &      -0.000         &      -0.001         &      -0.001         &      -0.000         \\
                    &     (0.000)         &     (0.000)         &     (0.000)         &     (0.002)         &     (0.002)         &     (0.002)         &     (0.002)         \\
Gender              &       0.030\sym{***}&       0.035\sym{***}&       0.034\sym{***}&       0.225\sym{***}&       0.211\sym{***}&       0.206\sym{***}&       0.218\sym{***}\\
                    &     (0.004)         &     (0.006)         &     (0.005)         &     (0.049)         &     (0.049)         &     (0.049)         &     (0.050)         \\
Level of Education  &       0.005\sym{***}&       0.007\sym{***}&       0.005\sym{**} &       0.025\sym{*}  &       0.043\sym{***}&       0.044\sym{***}&       0.036\sym{**} \\
                    &     (0.001)         &     (0.002)         &     (0.002)         &     (0.015)         &     (0.015)         &     (0.015)         &     (0.015)         \\
Treatment Condition: AI&       0.045\sym{***}&                     &                     &                     &                     &                     &                     \\
                    &     (0.003)         &                     &                     &                     &                     &                     &                     \\
Treatment Condition: High Confidence&       0.056\sym{***}&       0.017\sym{***}&       0.016\sym{***}&       0.093\sym{***}&       0.092\sym{***}&       0.093\sym{***}&       0.092\sym{***}\\
                    &     (0.003)         &     (0.004)         &     (0.004)         &     (0.034)         &     (0.034)         &     (0.034)         &     (0.034)         \\
AI Familiarity      &                     &                     &                     &       1.928\sym{***}&                     &                     &                     \\
                    &                     &                     &                     &     (0.388)         &                     &                     &                     \\
AI Familiarity Squared&                     &                     &                     &      -1.788\sym{***}&                     &                     &                     \\
                    &                     &                     &                     &     (0.463)         &                     &                     &                     \\
AI Knowledge        &                     &                     &                     &                     &       0.954\sym{**} &                     &                     \\
                    &                     &                     &                     &                     &     (0.412)         &                     &                     \\
AI Knowledge Squared&                     &                     &                     &                     &      -1.366\sym{***}&                     &                     \\
                    &                     &                     &                     &                     &     (0.530)         &                     &                     \\
AI Experience       &                     &                     &                     &                     &                     &       0.923\sym{***}&                     \\
                    &                     &                     &                     &                     &                     &     (0.280)         &                     \\
AI Experience Squared&                     &                     &                     &                     &                     &      -1.134\sym{***}&                     \\
                    &                     &                     &                     &                     &                     &     (0.300)         &                     \\
AI Background Index &                     &                     &                     &                     &                     &                     &       1.289\sym{***}\\
                    &                     &                     &                     &                     &                     &                     &     (0.483)         \\
AI Background Index Squared&                     &                     &                     &                     &                     &                     &      -1.547\sym{**} \\
                    &                     &                     &                     &                     &                     &                     &     (0.653)         \\
Normalized AI Sentiment&                     &                     &                     &      -0.349\sym{**} &      -0.203         &      -0.223         &      -0.274\sym{*}  \\
                    &                     &                     &                     &     (0.158)         &     (0.159)         &     (0.160)         &     (0.161)         \\
Constant            &       0.147\sym{***}&       0.219\sym{***}&       0.196\sym{***}&      -1.588\sym{***}&      -1.554\sym{***}&      -1.579\sym{***}&      -1.615\sym{***}\\
                    &     (0.011)         &     (0.015)         &     (0.015)         &     (0.167)         &     (0.174)         &     (0.171)         &     (0.180)         \\
\hline
Observations        &       90160         &       39903         &       40255         &       26214         &       26214         &       26214         &       26214         \\
\(R^{2}\)           &       0.020         &       0.015         &       0.014         &                     &                     &                     &                     \\
Pseudo \(R^{2}\)    &                     &                     &                     &       0.010         &       0.007         &       0.008         &       0.007         \\
Log Likelihood      &  -44495.143         &  -21683.412         &  -21555.620         &  -14202.630         &  -14237.277         &  -14229.660         &  -14236.006         \\
F                   &      88.738         &      26.239         &      27.421         &                     &                     &                     &                     \\
    \end{tabular}%
  }
  \caption{Analysis of Respondent Switching}
  \label{AnalysisOfRespondentSwitching}
  \small Notes: Standard errors clustered by respondent in parentheses. *p<0.10; **p<0.05; ***p<0.01.
\end{sidewaystable}

The results show an interaction between trust, confidence, knowledge, experience, and familiarity with AI, consistent with our theoretical expectations. There is a significant relationship between overall background in AI (hypothesis 1), level of trust in AI (hypothesis 2), treatment confidence level (hypothesis 3), practice round accuracy (hypothesis 4), and rate of switching at the $p<0.05$ level and above. We explore each of these results further in the subsequent sections.

We further test hypothesis 1 regarding how overall prior exposure to AI should influence automation bias in models 4-5 of Table~\ref{AnalysisOfRespondentSwitching}. We switch to a logit model here in part to highlight the consistency of the results across different model specifications, but the appendix shows that the model specification does not impact the results. We restrict the universe of cases to just those respondents that receive the AI treatment condition and evaluate the impact of our \textit{AI Background Index} variable, which includes the knowledge, experience, and familiarity sub-indices which we also examine below. Given our theoretical prediction that the relationship between overall background in AI and answer switching by respondents will be non-linear in hypothesis 1, we square the \textit{AI Background Index} variable.

\begin{figure}[H]
    \centering
    \includegraphics[width = 0.7\linewidth]{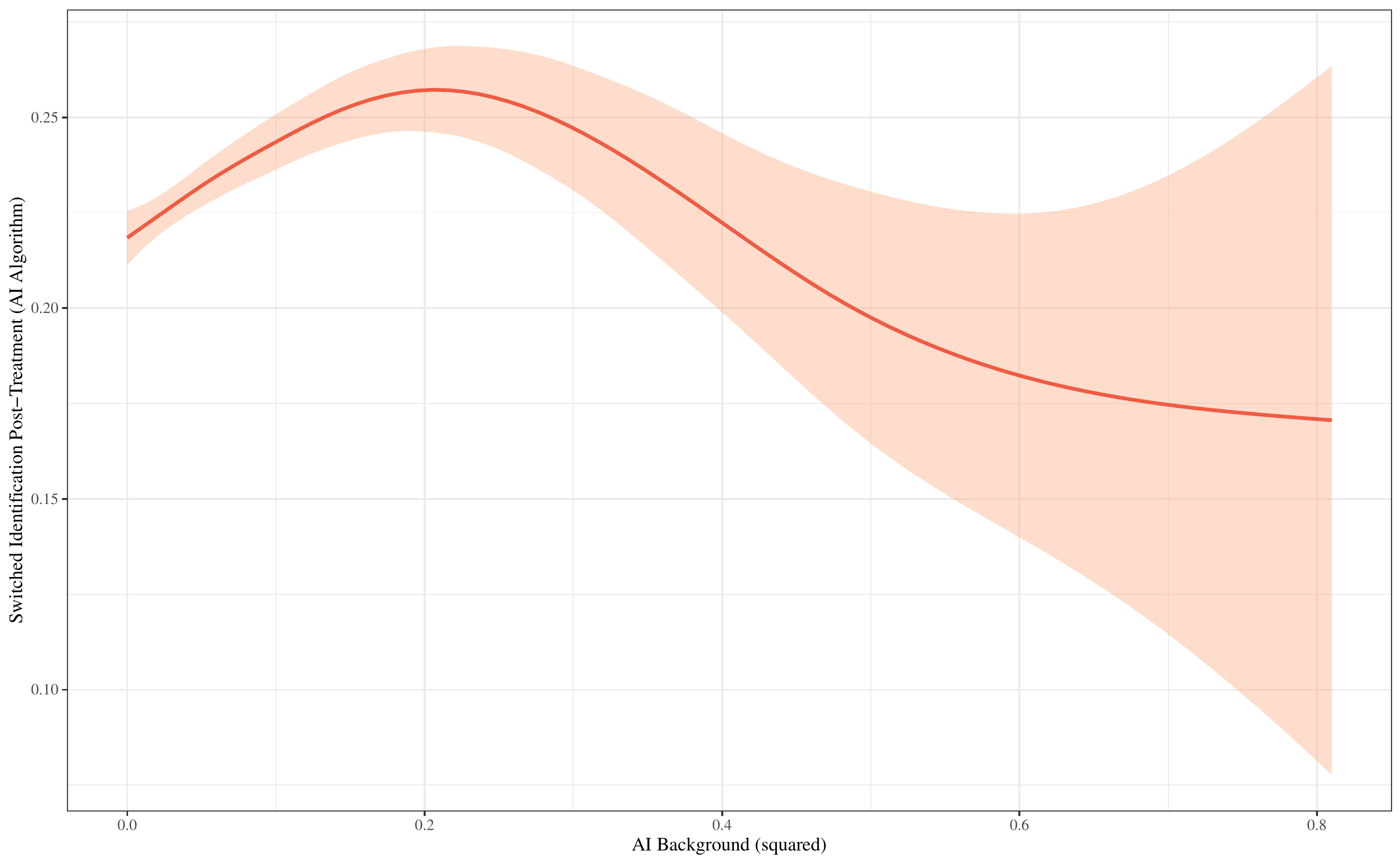}
    \caption{Rate of switching when treatment is an AI algorithm, given AI Background}
    \label{fig:AIBackgroundCurve}
\end{figure}

The statistical results show strong and significant interaction effects and highlight the non-linearity of the effects. We begin evaluating the substantive effects by switching back from regression to comparing mean values of switching after receiving an AI algorithm treatment, as values of the \textit{AI Background Index} increase. Figure~\ref{fig:AIBackgroundCurve} plots the rate of switching. The predicted non-linear relationship is evident, supporting the theory laid out in Figure~\ref{fig:reliancefig}. Reliance on AI (in this case, the rate of switching when receiving an AI treatment) initially increases steadily at the lowest combined level of AI familiarity, knowledge, and experience, peaking when \textit{AI Background Index} is at a middling level, around the mean ($0.224$) and median values ($0.196$) of \textit{AI Background Index}, before it decreases and levels out as \textit{AI Background Index} increases past the 3rd quantile mark of around $0.313$ and trends towards the maximum of $0.900$.

\begin{figure}[H]
    \centering
    \begin{tabular}{cc}
    \adjustbox{valign=m}{\subfloat{%
          \includegraphics[width=0.65\linewidth,height=8.1cm]{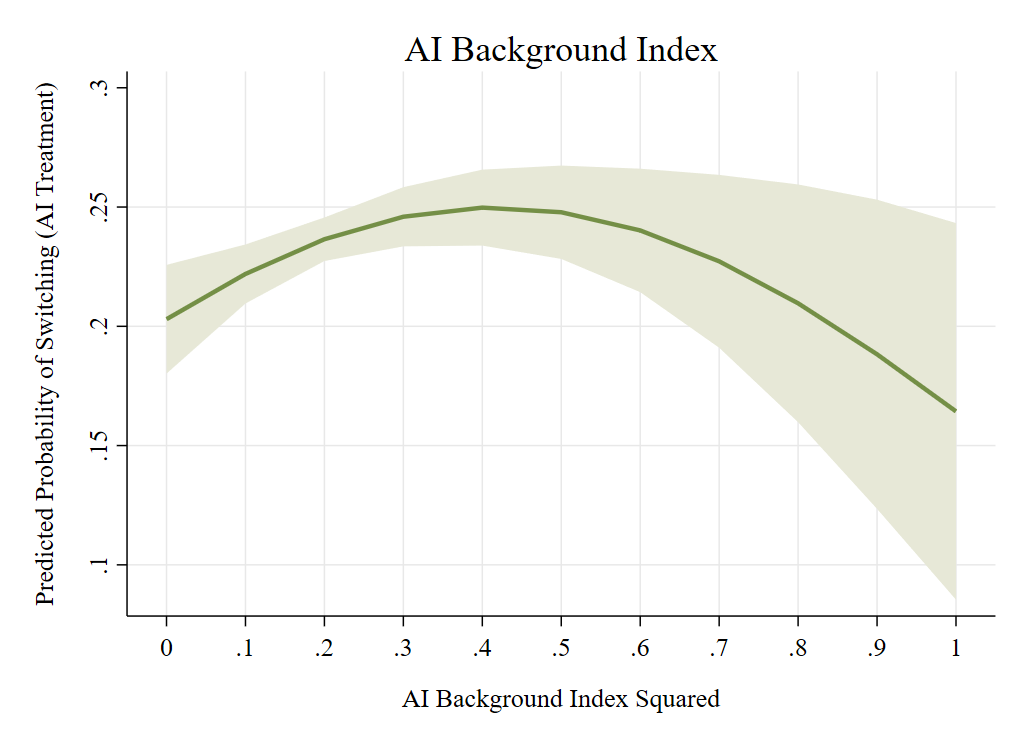}}}
    &      
    \adjustbox{valign=m}{\begin{tabular}{@{}c@{}c@{}}
    \subfloat{%
          \includegraphics[width=.4\linewidth]{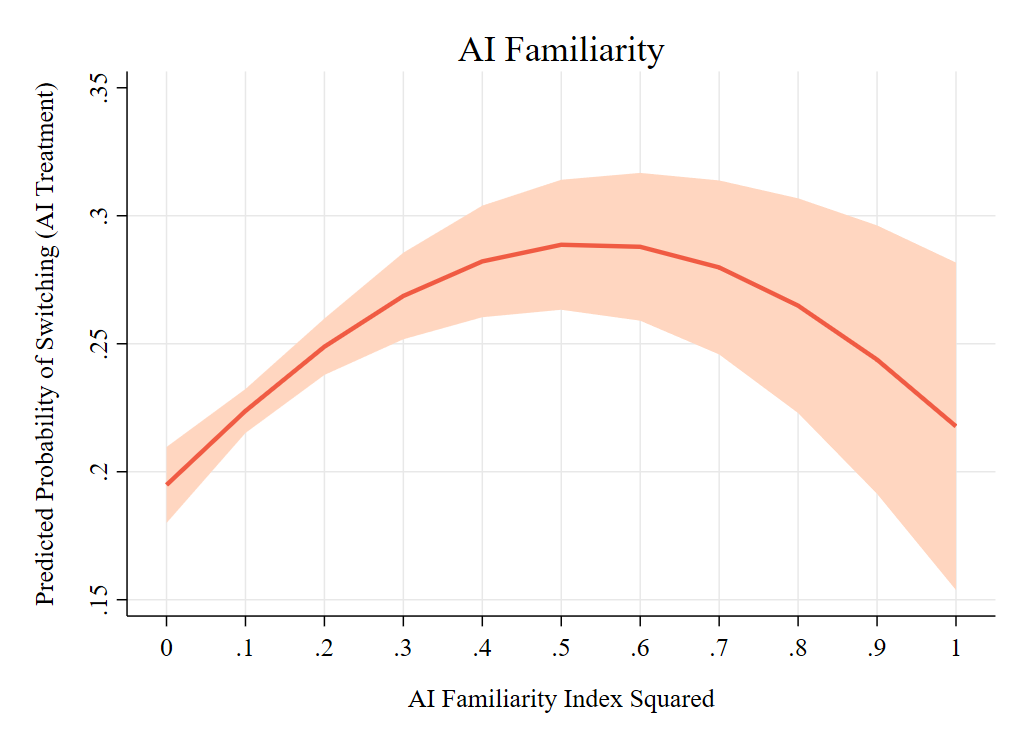}} \\
    \subfloat{%
          \includegraphics[width=.4\linewidth]{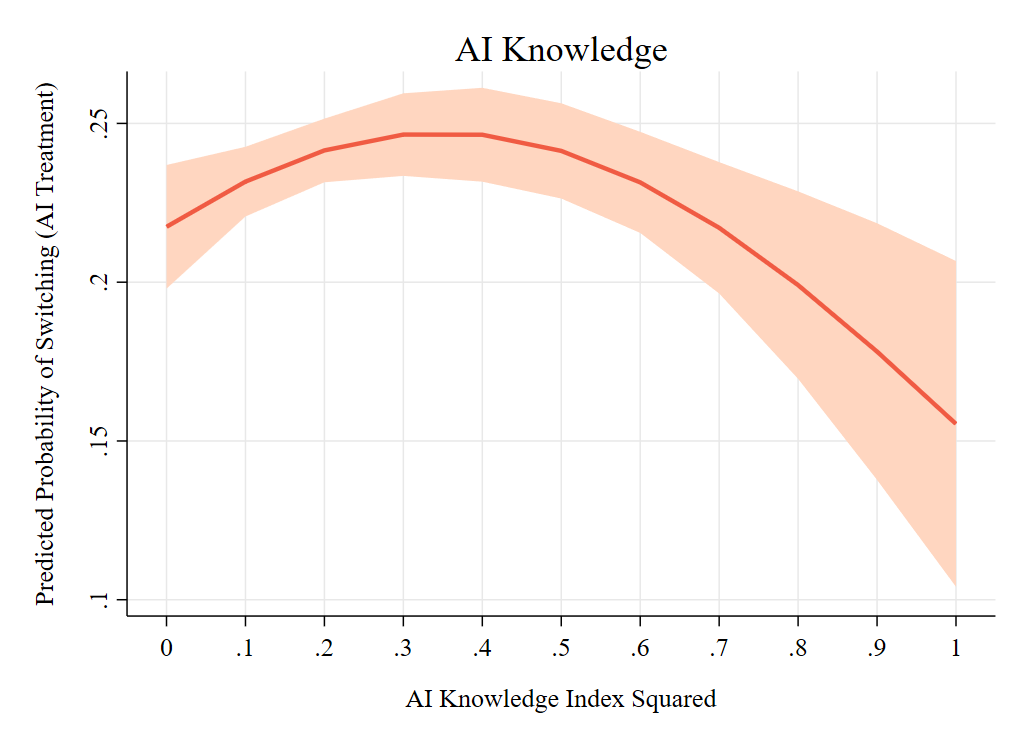}} \\
    \subfloat{%
          \includegraphics[width=.4\linewidth]{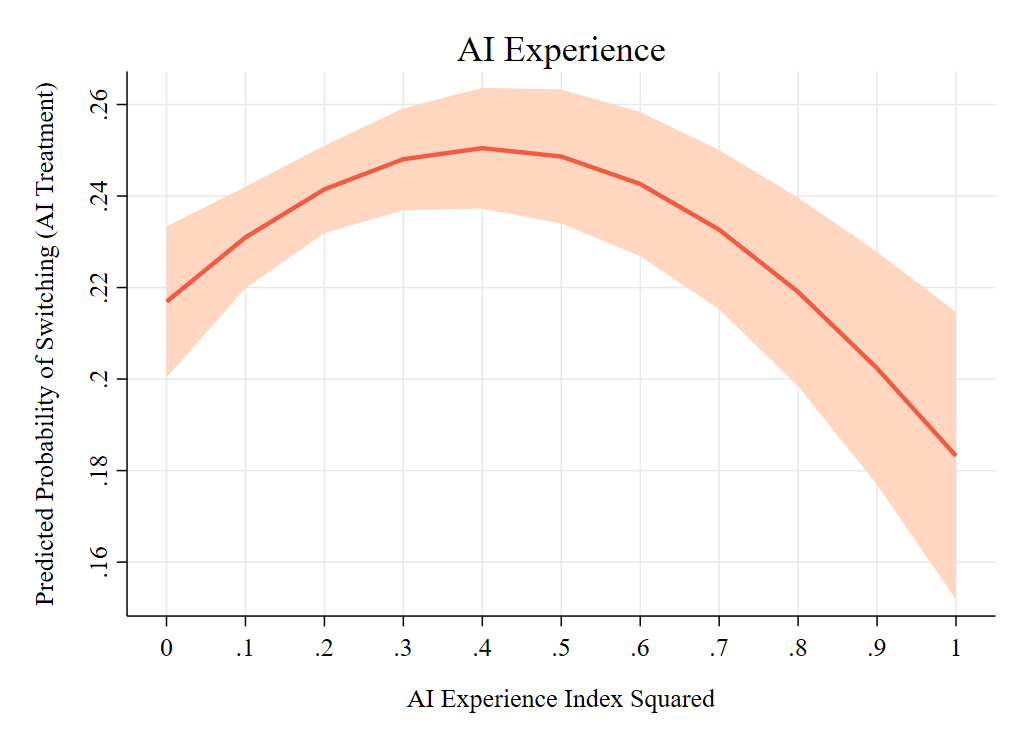}}
    \end{tabular}}
    \end{tabular}
    \caption{Predicted Probability of Switching: AI Background and Index Components}\label{fig:ProportionalQuadChart}
 \end{figure}

We then evaluate the substantive effects of each of the sub-indices that together constitute the \textit{AI Background Index}, and the background index itself, in Figure~\ref{fig:ProportionalQuadChart}, based on the additional models in Table~\ref{AnalysisOfRespondentSwitching}. We generate predicted probabilities for switching across each of the values of the \textit{Overall AI Familiarity}, \textit{Overall AI Knowledge}, \textit{Overall AI Experience}, and \textit{AI Background Index} variables, holding other variables at their means, and plot the results. The results curves look slightly different than in Figure~\ref{fig:AIBackgroundCurve} above due to the addition of control variables in the regression framework, the addition of country fixed effects, and clustered standard errors. Their consistency, however, demonstrates a clear pattern of results that supports our hypothesis.

The probability of switching in the AI condition is 19\% when \textit{Overall AI Familiarity} is lowest. Consistent with the Dunning-Kruger effect \citep{SanchezDunning2018}, as perceived AI familiarity grows, respondents become more likely to switch their answers in response to information provided by an AI decision aid, peaking at 29\%. As \textit{Overall AI Familiarity} gets to the highest level, though, the probability of switching goes down again to 22\%. 

We generally see similar effects for \textit{Overall AI Knowledge} (starts at 22\%, peaks at 25\%, ends at 16\%) and \textit{Overall AI Experience} (starts at 20\%, peaks at 25\%, ends at 18\%), with the \textit{Overall AI Knowledge} curve being slightly steeper, which is expected considering the much higher level of difficulty of the AI knowledge questions. The overall index starts at 20\% when \textit{AI Background Index} is lowest, peaks at 25\% as background rises, and drops to 16\% when \textit{AI Background Index} is at its highest. The results provide strong initial support for hypothesis 1.

\subsection{Attitudinal: trust and confidence}

We now look at the impact of the level of confidence and trust the respondents have in the decision-making aid and themselves on the rate of post-treatment identification switching, the main dependent variable. As outlined earlier, we measure confidence in the system with our reliability treatment condition, \textit{Treatment Confidence}). As a reminder, we also measure self-confidence as the percentage of practice rounds accurately identified (\textit{Total Number of Practice Rounds Correct}). Respondents received live feedback on their accuracy in these initial five practice rounds and, thus, had an approximate sense of how well they completed the task. We also separately generated a proxy for trust in AI using a battery of questions that generated a score for respondents (\textit{Trust in AI}) based on how negatively or positively they reacted to statements about AI.


Figure~\ref{fig:LikertBeliefs} below shows the percentage of respondents that agreed or disagreed with each statement used to generate the AI Trust Index. 
\begin{figure}
    \centering
    \includegraphics[trim={0 0 0 1cm},clip, width = 1.0 \linewidth]{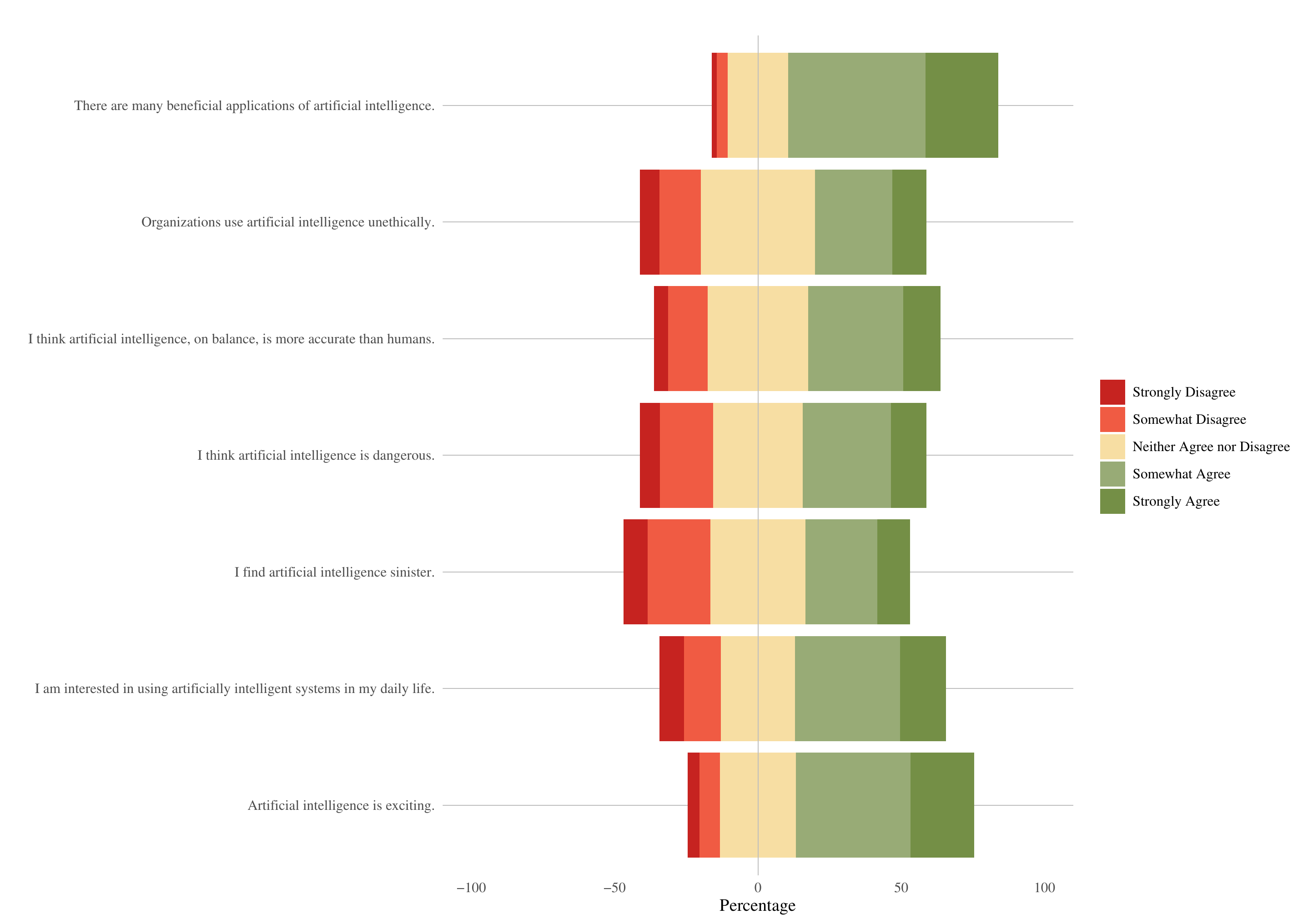}
    \caption{Responses to AI Beliefs Battery}
    \label{fig:LikertBeliefs}
\end{figure}

Overall, most respondents have some degree of openness to artificial intelligence. A majority found artificial intelligence to be exciting and to have many potential benefits, with 62\% and 73\%, respectively, of the sample either responding with "agree" or "strongly agree", while over half (53\%) of respondents indicated an interest in using "artificially intelligent systems." More pertinent for this study, about 46\% even agreed that AI was, on balance, more accurate than humans. There was some hesitation and worry over the use of AI. However, 43\% responded that they believed AI to be "dangerous," 39\% found that organizations use AI unethically, and 36\% even argued that they find AI "sinister." Together, the responses sketch a general picture of how open and trusting individuals were of AI technologies and allowed us to generate the \textit{Trust in AI} index.

\begin{figure}[H]
    \centering
    \includegraphics[width = \linewidth]{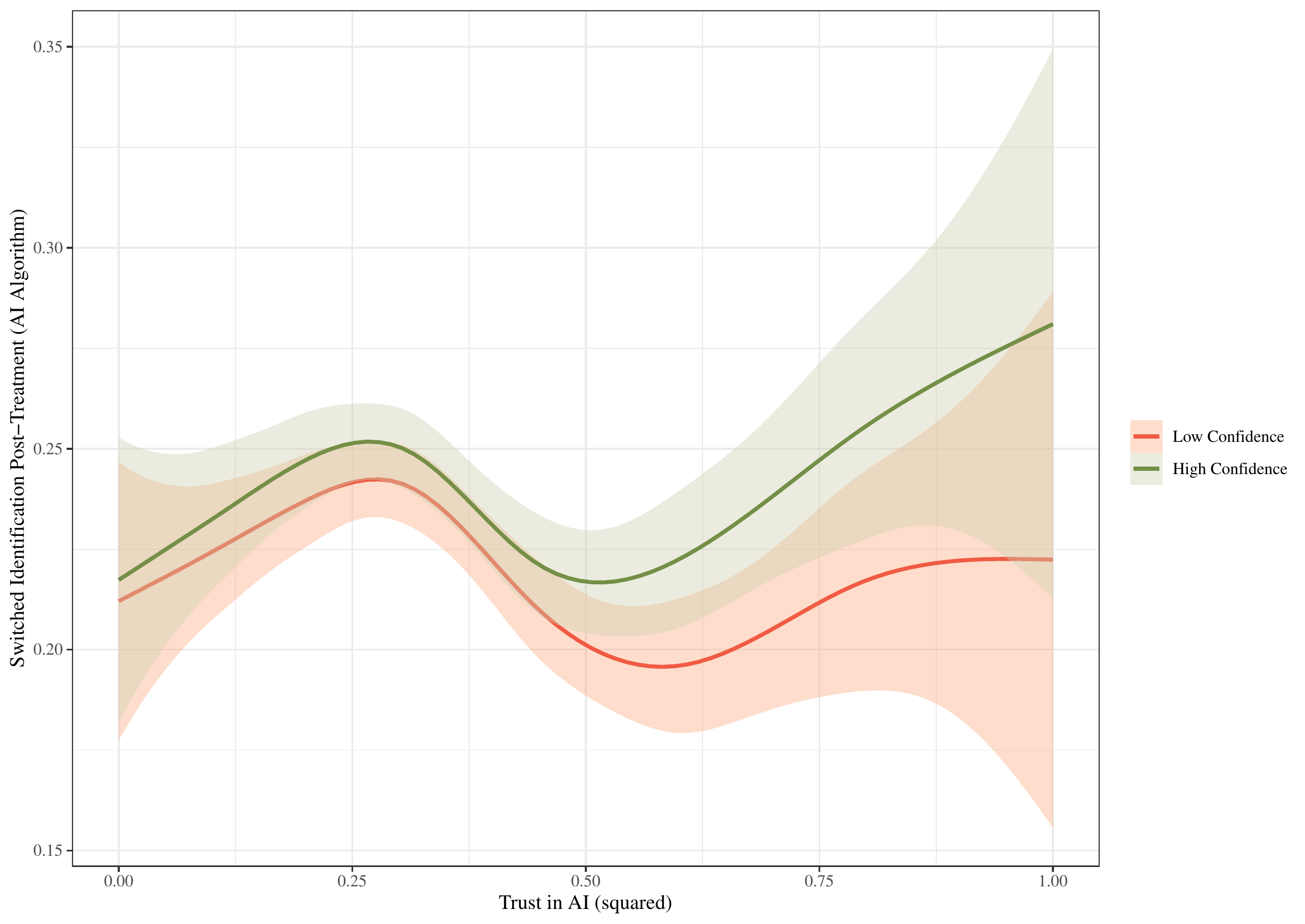}
    \caption{Average Switching Given an AI Treatment Across AI Trust Levels}
    \label{fig:SwitchedAISentimentConfidence}
\end{figure}

To further test hypothesis 2, we assess the relationship between the intersection of a respondent's trust and confidence in AI and their likelihood to listen to an AI algorithm-based decision aid or be more biased towards automation. Figure~\ref{fig:SwitchedAISentimentConfidence} plots respondents' rate of switching their answers when shown an AI algorithm treatment, given their overall level of trust in AI-based technology and systems. Generally, the switching rate for both high and low-confidence AI algorithm treatments showed a similar, non-linear trend as trust in AI increased but was overall higher for high-confidence treatments. Similar to the combined effects of knowledge, experience, and familiarity with AI, individuals are more prone to automation aversion at the lowest levels of trust.

Confidence in the specific AI algorithm mattered less to respondents with lower general levels of trust in AI. For those respondents with an \textit{Trust in AI} index score less than the median ($0.571$) (or mean $0.573$), their personal perception of the technology outweighed the influence of the experimental information on the level of testing or training the AI algorithm received. However, Respondents with a greater than mean or median level of trust in AI were substantially more likely to switch answers if the AI algorithm was established to have a higher degree of accuracy by already having completed "extensive testing and training." This suggests that once the barrier of trust in the technology is overcome, confidence --- the testing, evaluation, and the expected accuracy of the system --- becomes a more central factor in willingness to rely on the technology.

The above also shows evidence for hypothesis 3. In general, when the decision aid for the respondent is described in terms that elicit higher degrees of confidence, respondents are significantly more likely to change their answers. The findings hold whether respondents received the human or AI experimental treatments.

\begin{figure}[H]
    \centering
    \begin{tabular}{cc}
    \adjustbox{valign=b}{\subfloat{%
          \includegraphics[width=0.5\linewidth]{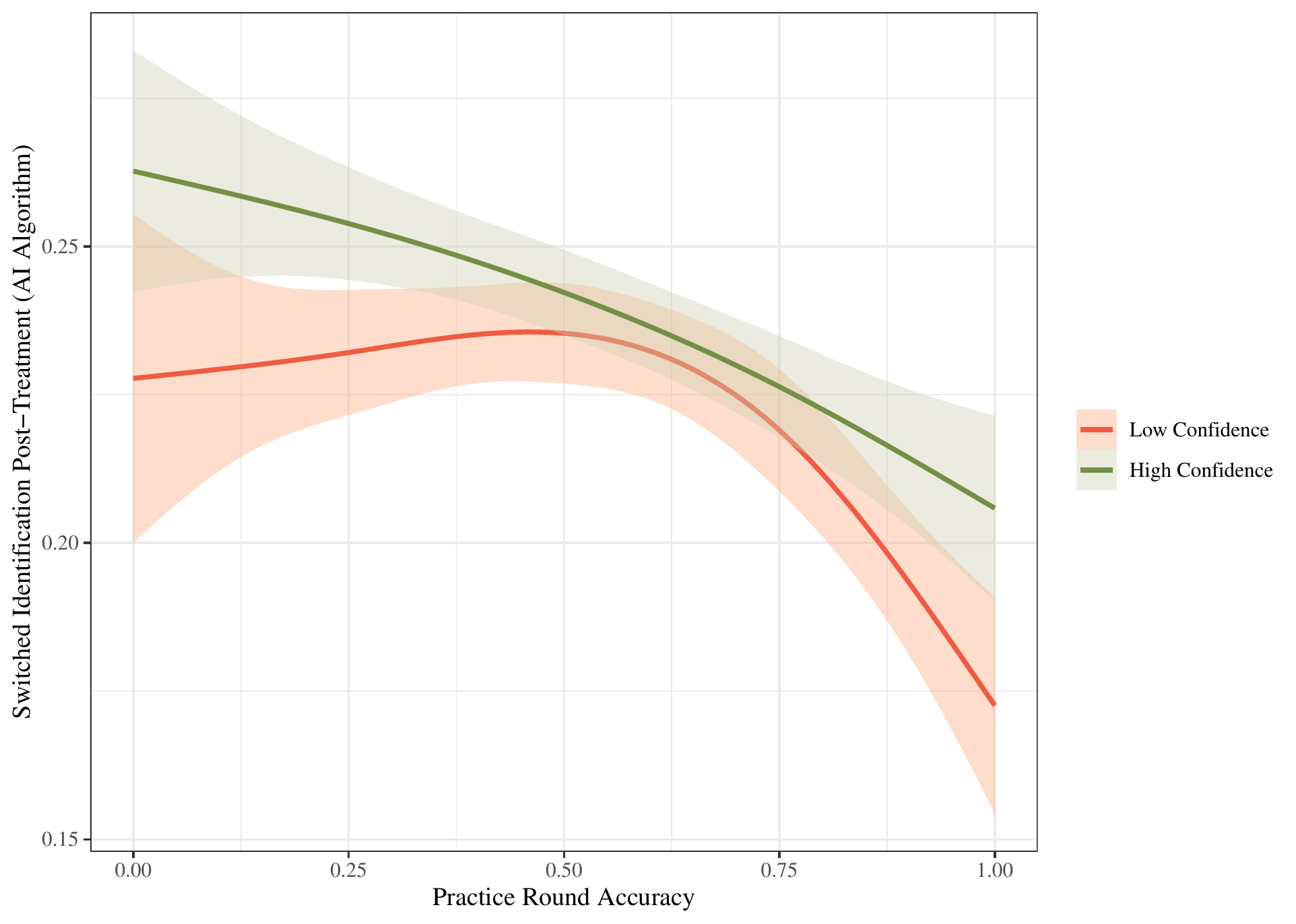}}}
    &      
    \adjustbox{valign=b}{\begin{tabular}{@{}c@{}}
    \subfloat{%
          \includegraphics[width=.5\linewidth]{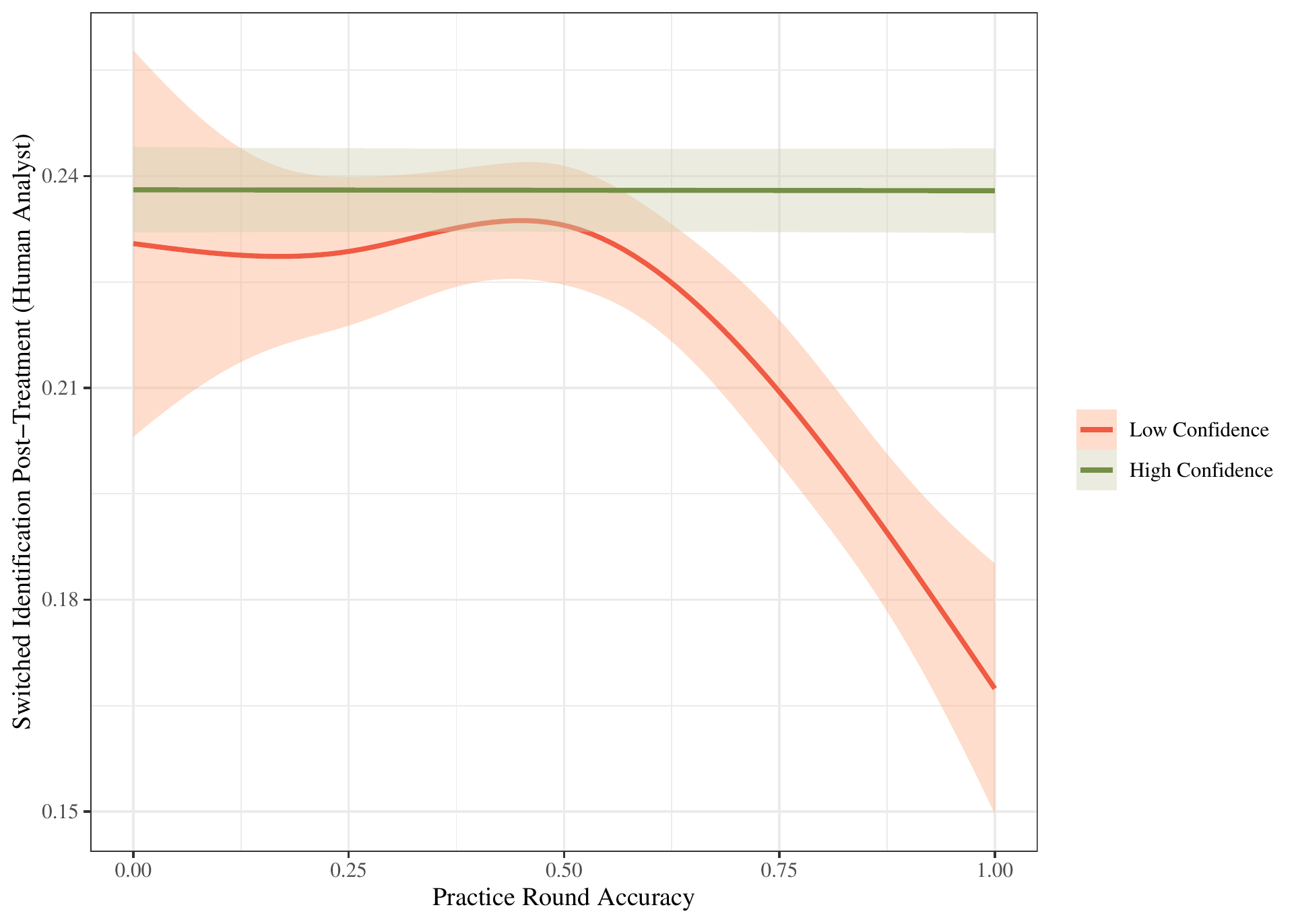}}
    \end{tabular}}
    \end{tabular}
    \caption{Rate of Switching Given Treatment Type and Accuracy in Practice Rounds}\label{fig:SelfConfidenceFigure}
 \end{figure}

Finally, we find support for hypothesis 4. Figure~\ref{fig:SelfConfidenceFigure} shows the rate of a respondent switching answers after being shown each treatment type, given their practice round accuracy. The higher the score of a respondent on the practice round identification tasks, the less likely they were to switch their answers in the "real" identification rounds for most treatment scenarios. Self-confidence, derived from the practice rounds, serves as perceived task competence---the higher the level of self-confidence, the lower the probability that respondents will rely on decision aids. 

For those respondents that got no practice round identifications correct, they switched their answers most when presented with an AI treatment, having changed their answers over 25\% of the time, whereas those respondents who achieved a perfect score in the practice round switched their answers only 18\% of the time. While respondents that had low practice round accuracy switched their answers at a relatively constant rate for AI treatments, regardless of how they were described, those that had the highest practice round accuracy changed their answers for high-confidence AI treatments about 20\% of the time, but noticeably less frequently for low-confidence AI treatments (16\%). The one exception involves the high-confidence human analyst condition; regardless of the degree of practice round accuracy, the switching rate remained constant, at around 24\% of the time. Excluding that exception, this shows that there is more variation in the confidence respondents have in advice from humans than confidence respondents have in advice from AI, especially when one's own perceived competency and self-confidence are taken into account.

\section{Conclusion}

What influences the adoption of new technologies is a crucial question for international politics. The use of artificial intelligence by governments, in particular, involves questions of trust, confidence, and human agency that are fundamental to our understanding of international relations, particularly as AI increasingly shapes political decisions. As the integration of AI into militaries worldwide continues, how individuals and organizations make decisions about adopting AI will become even more critical. Research on automation bias suggests that humans can often be overconfident in AI, whereas research on algorithm aversion shows that, as the stakes of a decision rise, humans because more cautious about trusting algorithms. We theorize about the relationship between background knowledge about AI, trust in AI, and how these interact with other factors to influence the probability of automation bias in a critical international political context. We test our hypotheses across a sample of 9000 respondents in 9 countries with extensive but varying levels of AI industries and military investments. The results strongly support the theory, especially concerning AI background knowledge. A version of the Dunning Kruger effect appears to be at play, whereby those with the lowest level of experience with AI are slightly more likely to be algorithm-averse, then automation bias occurs at lower levels of knowledge before leveling off as a respondent's AI background reaches the highest levels.
 
We find that there is a nonlinear relationship between how likely an individual will rely on an automated, AI-based decision aid in the international security context and their overall exposure to AI, where exposure encompasses the extent to which an individual has a general familiarity and awareness of the technology, concrete knowledge of how the technology works, and firsthand experience using the technology. The interaction of experience, knowledge, and familiarity drives responses to decision aids, where at the lowest levels of these factors, automation aversion is most likely, and at average levels, automation bias is most likely. 

Only at the highest levels of overall exposure to AI are individuals more balanced in whether they rely on the AI decision aid for military aircraft identification, and we see greater accuracy in task completion, particularly amongst those respondents who had low levels of self-confidence in their ability to complete the task. Appendix figure~\ref{fig:RespondentAccuracy} details these findings, illustrating how respondents with the lowest levels of self-confidence in the ability to complete the task were more likely to switch initially incorrect answers when suggested a correction by an algorithm-based decision aid, with greater accuracy as their background in AI increased, from just around 50\% accuracy to nearly 70\% accuracy.
 
We also find that trust and confidence play a significant role in whether an individual will act on recommendations made by the automated decision aid. We find that the more positive someone's attitudes towards AI technology are, the more trusting they are in that technology, and the more likely the individual is to trust a recommendation of an AI-enabled system over that of a human analyst. Similarly, we find that the greater the described confidence and accuracy in the system, the more likely a respondent is to follow the system's decision suggestions. However, we find that this is often tempered by a respondent's self-confidence---the more accurate a respondent was described to be in the initial practice segment, the less likely they are to second-guess their decisions and over-rely on a decision aid, regardless of whether it was a human analyst or an AI algorithm making the suggestion.
 
There are several limitations to this study that could inform future research. First, this study focused on the military domain because most previous automation bias experiments did not. Future research could incorporate a non-military scenario and a military scenario to allow for the direct comparability of results. Second, rather than focusing solely on public views, future research could explore the perspectives of elites or those directly involved in national security and military efforts in various countries. Third, future research could look at additional tasks in the military domain to see how automation bias and algorithm aversion vary across the type of task.
 
Overall, these results demonstrate that the way people think about AI and the features of any specific AI-enabled system influences how the system is used in an international security context, including rates of automation bias and algorithmic aversion. It also builds on existing behavioral science research on cognitive biases and receptivity to using automated and increasingly computerized tools. These findings not only contribute new theory but also illustrate the importance of understanding trust and confidence in AI adoption and how AI will broadly influence international relations.


\bibliographystyle{apalike}
\bibliography{paper}


\newpage
\appendix

\counterwithin{figure}{section}
\counterwithin{table}{section}

\section{Online Appendix}

\subsection{Additional Figures and Tables}

\begin{figure}[H]
    \centering
    \includegraphics[trim={1cm 2cm 3cm 1cm},clip, width = \textwidth]{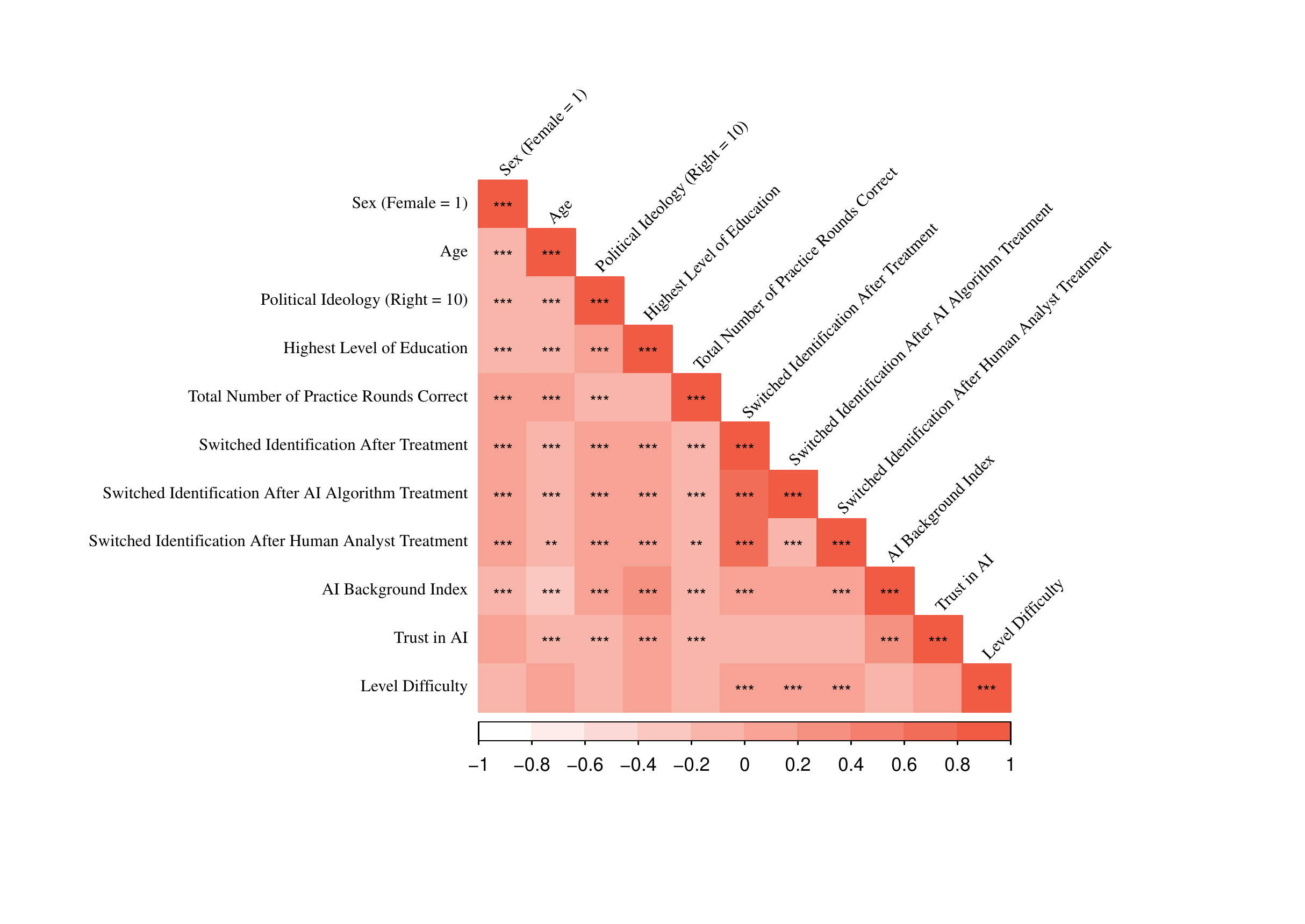}
    \caption{Correlations and their significance between select variables.}
    \label{fig:CorrelationPlot}
\end{figure}

\begin{figure}[H]
    \centering
    \includegraphics{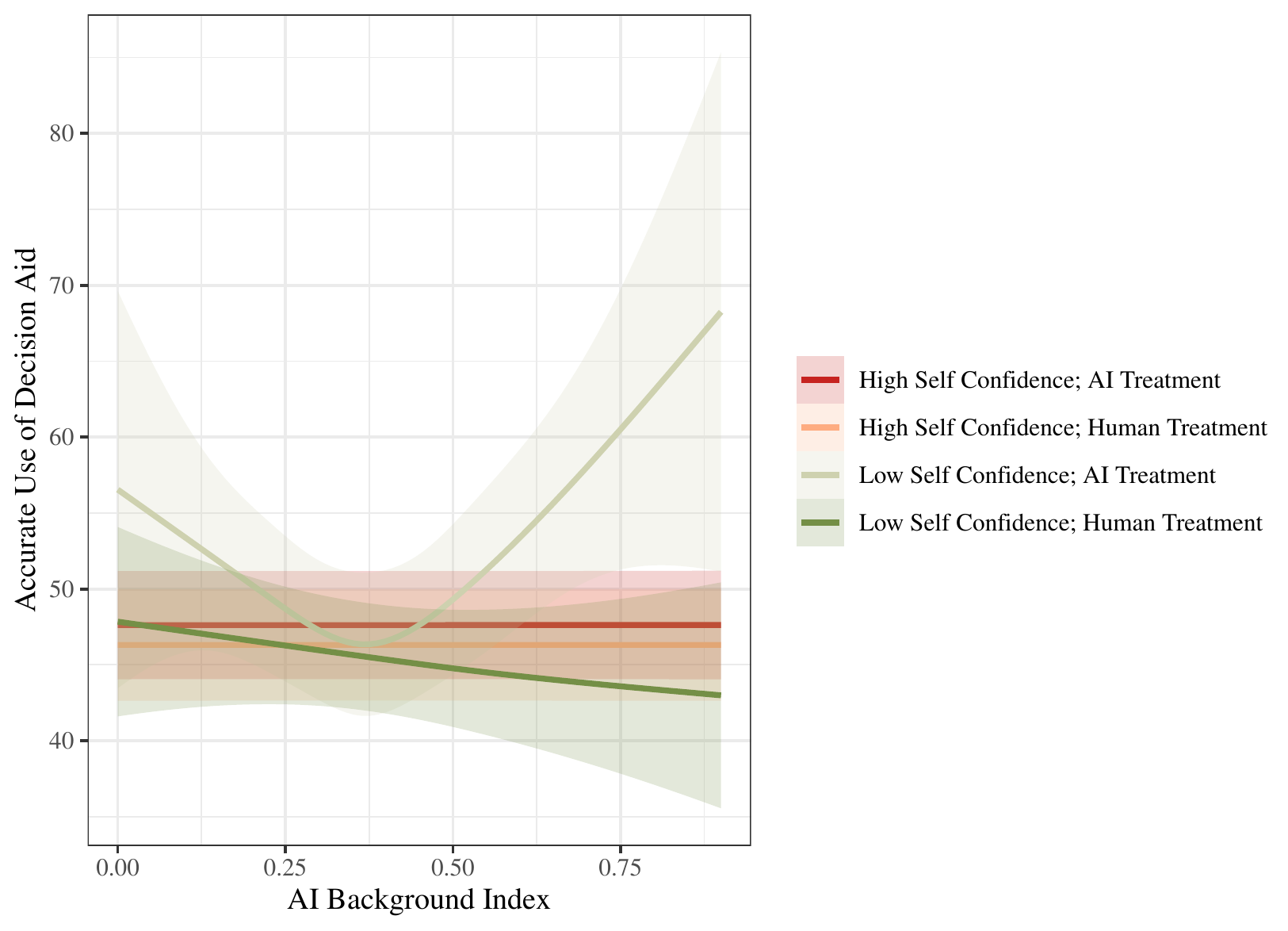}
    \caption{Accuracy of respondents in aircraft identification by their background in AI and degree of self-confidence. Respondents were coded as having high self-confidence if they obtained a score of 60\% or greater in the practice rounds. Accurate use of a decision aid was considered when an original identification was incorrect, the treatment suggestion was correct, and the respondent switched to the correct answer (scored as 100). Inaccurate use of a decision aid was considered when an original identification was correct, the treatment suggestion was incorrect, and the respondent switched to the incorrect answer (scored as 0).}
    \label{fig:RespondentAccuracy}
\end{figure}

\begin{figure}[H]
    \centering
    \includegraphics[width = \linewidth]{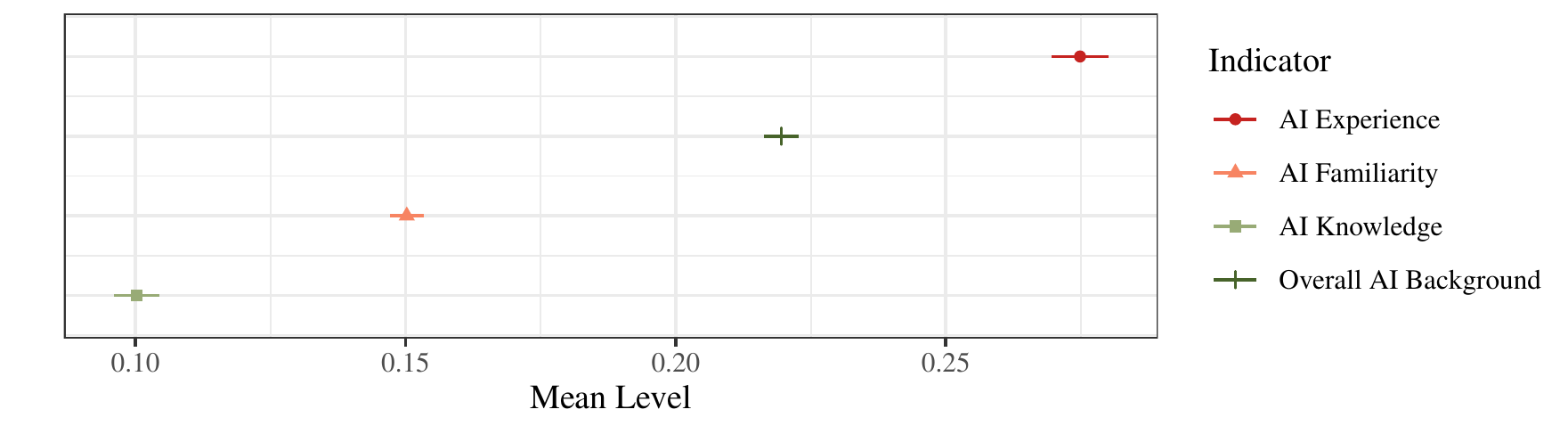}
    \caption{Mean Values of AI Background and Sub-indices}
    \label{fig:AIBackgroundDetails}
\end{figure}

\newpage
\clearpage
\subsection*{Example Test Scenarios}

\begin{figure}[H]
    \centering
    \includegraphics[width = 0.7\textwidth]{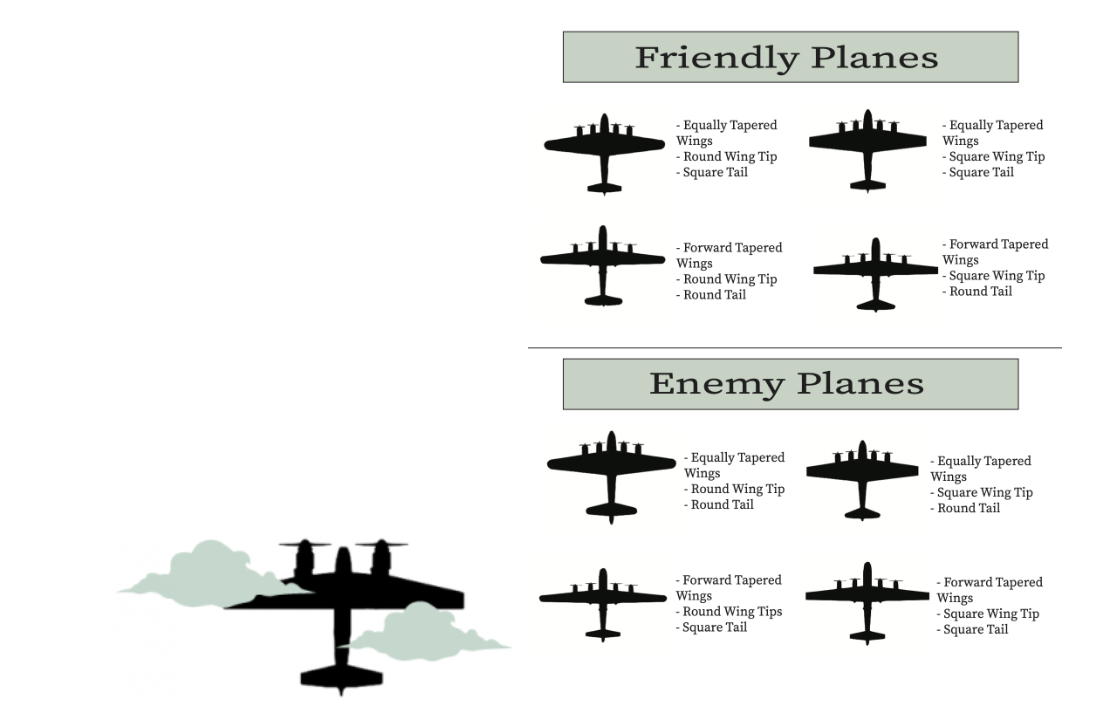}
    \caption{An example of an airplane identification scenario at the easier difficulty level.}
    \label{fig:Example Easy}
\end{figure}

\begin{figure}[H]
    \centering
    \includegraphics[width = 0.7\textwidth]{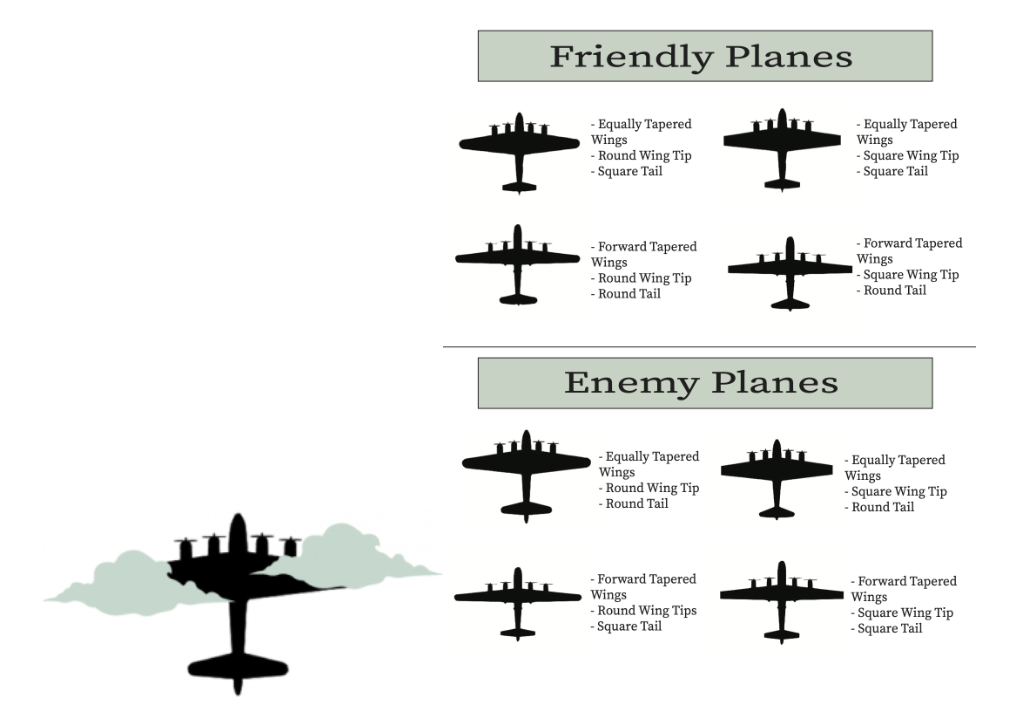}
    \caption{An example of an airplane identification scenario at the more challenging difficulty level.}
    \label{fig:Example Difficult}
\end{figure}

\newpage

\subsection*{Trust in AI Battery}

Please indicate the extent to which you agree or disagree with each of the following statements:

\begin{itemize}
    \item "Organizations use artificial intelligence unethically."
    \item "I am interested in using artificially intelligent systems in my daily life."
    \item "I find artificial intelligence sinister."
    \item "I think artificial intelligence is dangerous."
    \item "Artificial intelligence is exciting."
    \item "I think artificial intelligence, on balance, is more accurate than humans."
    \item "There are many beneficial applications of artificial intelligence."
\end{itemize}

Answer options:
\begin{itemize}
    \item Strongly disagree
    \item Somewhat disagree
    \item Neither agree nor disagree
    \item Somewhat agree
    \item Strongly agree
    \item Don't know
\end{itemize}

\end{document}